\documentclass[%
reprint,
 amsmath,amssymb,
 aps,
 prl,
]{revtex4-2}
\usepackage{soul}
\usepackage{xcolor}
\usepackage{url}
\usepackage{graphicx}
\usepackage{dcolumn}
\usepackage{bm}
\usepackage{physics}
\usepackage{upgreek}
\allowdisplaybreaks

\begin{document}

\preprint{APS/123-QED}

\title{Generation of cold Rydberg atoms at submicron distances from an optical nanofiber}

\author{Krishnapriya Subramonian Rajasree$^1$}%
\author{Tridib Ray$^1$}%
\altaffiliation[Current address: Laboratoire Kastler Brossel, Sorbonne Universit\'{e}, CNRS,
ENS-Universit\'{e} PSL, Coll\`{e}ge de France, 4 place Jussieu, F-75005 Paris, France]{}
 \author{Kristoffer Karlsson$^1$}%
 \author{Jesse L. Everett$^1$}%
  \author{S\'{i}le Nic Chormaic$^{1,2}$}%
  \email{Sile.NicChormaic@oist.jp}%

\affiliation{
$^{1}$Light-Matter Interactions for Quantum Technologies Unit, Okinawa Institute of Science and Technology Graduate University, Onna, Okinawa 904-0495, Japan\\
$^{2}$Universit\'{e} Grenoble Alpes, CNRS, Grenoble INP, Institut N\'{e}el, 38000 Grenoble, France
}%


\date{\today}

\begin{abstract}
We report on a  controllable, hybrid quantum system consisting of cold Rydberg atoms and an optical nanofiber interface. Using a two-photon ladder-type excitation in $^{87}$Rb, we demonstrate both coherent and incoherent Rydberg excitation at submicron distances from the nanofiber surface. The 780 nm photon, near resonant to the $5S \rightarrow 5P$ transition, is mediated by the cooling laser, while the  482 nm light, near resonant to the $5P \rightarrow 29D$ transition, is mediated by the guided mode of the nanofiber. The population loss rate of the cold atom ensemble is used to measure the Rydberg population rate. A theoretical model is developed to interpret the results and link the population loss rate to the experimentally measured, effective Rabi frequency of the process. This work makes headway in the study of Rydberg atom-surface interactions at submicron distances and the use of cold Rydberg atoms for all-fibered quantum networks. 
\end{abstract}

\maketitle

In recent years, Rydberg atoms have emerged as leading candidates for neutral atom based quantum information processing \cite{Jaksch2000,Zoller2001,Saffman2010R,adams_rydberg_2019} and quantum simulations \cite{weimer2010,Antoine2016,Lukin2017}. The long-lived quantum states and the precisely tunable dipolar interaction, leading to blockade \cite{Gould2004}, can be used to prepare a mesoscopic atomic ensemble exhibiting quantum correlations and entanglement \cite{Antoine2010}. Such systems have already been used to demonstrate a quantum phase gate \cite{Saffman2010} and simulator \cite{Lukin2017} in free-space. Interfacing interacting Rydberg atoms with microfabricated components is a very attractive choice for building compact and scalable hybrid quantum devices. Coherent Rydberg excitation has been reported for an atom chip \cite{Druten2018},  a $\mu$m-sized vapor cell \cite{Pfau2010}, a hollow-core photonic crystal fiber \cite{Epple2014, Windpassinger2017} and has been proposed for a superconducting resonator \cite{Dumke2017}, with each platform having its own advantages and disadvantages. 

Here, we present an alternative, but highly viable, platform for atom-based quantum networks by interfacing cold Rydberg atoms with a single-mode optical nanofiber (ONF). To date, ground-state cold atom-ONF hybrid systems \cite{Nieddu2016,Rolston2017R} have shown tremendous potential for a new generation of quantum  devices. The small cross-section of the evanescent field from an ONF, as a result of the exponential radial decay from the fiber waist, leads to a large co-operativity \cite{Hakuta2005,Hakuta2006}, the key to many quantum optics experiments. The high intensity and field gradient can be used to optically trap atoms in a one-dimensional array \cite{Hakuta2004,PhysRevLett.104.203603}, thereby enabling the study of one dimensional, many-body physics, or can be exploited for quadrupole transition enhancement \cite{PhysRevA.97.013821}. Cooperative effects, such as the generation of a collective entangled state \cite{Laurat2019}, have been demonstrated in such a system using ground-state, neutral Cs atoms. In addition, atoms in the evanescent field region are intrinsically coupled to an optical bus in the form of the fiber-guided mode. This can lead to low-loss transfer of information to and from the interaction region \cite{Stourm_2019}, a prerequisite for Rydberg-based quantum repeaters in fiber-coupled cavities \cite{PhysRevA.85.042324}. 

Aside from endeavours to combine ONFs and ground-state, neutral atoms, work on Rydberg excitation next to ONFs has, to-date, been limited to theoretical proposals due to the difficulty in generating highly excited atom states within a few 100 nm of surfaces, e.g. dielectrics or metals, and the problem of induced electric fields - even at distances as large as $\sim$100 $\mu$m - by adsorption of atoms on  the surface \cite{PhysRevLett.116.133201,PhysRevA.81.063411}. In this work, we report on evanescent field assisted Rydberg excitation at submicron distances from the surface of an ONF, which is embedded in a $^{87}$Rb atomic ensemble in a magneto-optical trap (MOT).  A ladder-type, two-photon excitation scheme \cite{Adams2007} is used to excite the atoms to the Rydberg state and a trap loss method \cite{Walker2008} is used to probe the Rydberg excitation.  We implement a rate equation model \cite{Sadiq2013} to determine the rate of population transfer to the Rydberg state. Both coherent two-photon excitation and incoherent two-step excitation is demonstrated. A density matrix based model is developed for the three-level ladder-type system interacting with the evanescent field of the ONF. The developed model  explains the main features of our experimental results.

\begin{figure}[t]
\center
\includegraphics[width=8.6 cm]{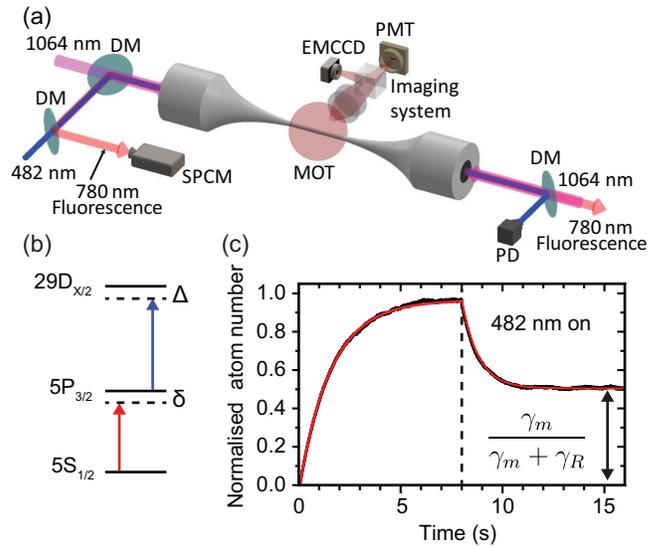} 
\caption{ (a) Experimental setup. DM: Dichroic Mirror, PD: Photodiode, PMT: Photomultiplier tube, EMCCD: Electron multiplying charge-coupled device, SPCM: Single-photon counting module, Imaging system: Lens and beam-splitter combination.  (b) Simplified $^{87}$Rb energy diagram. Atoms trapped in the MOT are excited to the $n=29$ level by a two-photon process involving the cooling laser at 780 nm and  482 nm light in the evanescent field of the ONF. (c) Typical loading of the MOT to a steady-state for 8 s and subsequent decay to a new equilibrium for 8 s after the 482 nm laser has been switched on.}
\label{Fig:setup}
\end{figure}

The experiment consists of an ONF, with a waist diameter of $\sim$400 nm, that is single-mode for 780 nm, embedded in a cold  atomic ensemble of $^{87}$Rb.  A schematic of the experimental setup is given in Fig. \ref{Fig:setup}(a).  The ONF is fabricated by exponentially tapering a section of SM800-125 fiber (cut-off wavelength 697 nm)  using a H:O flame-brushing technique \cite{doi:10.1063/1.4901098}. To guarantee a low-loss ONF in the ultrahigh vacuum (UHV) chamber, we ensure that the tapering process is adiabatic and that the  fiber itself is very clean. During the experiment, 100 $\mu W$ of 1064 nm light is passed through the ONF from each side to keep the fiber hot and avoid atom deposition on the ONF \cite{doi:10.1063/1.5027743}. This light has the added advantage of attracting atoms toward the ONF surface, thereby increasing the number of atoms interacting with the evanescent field.

 The $^{87}$Rb atoms are cooled to $\sim$120 $\mu$K using a standard MOT configuration of three pairs of retro-reflected cooling and repump beams. The 780 nm cooling  beam is 14 MHz red-detuned from the 5$S_{1/2}(F= 2) \rightarrow$ 5$P_{3/2}(F^\prime= 3)$ transition and the repump is on resonance with the 5$S_{1/2}(F= 1)$ {$\rightarrow$} 5$P_{3/2}(F^\prime= 2)$ transition. The total powers in the  cooling and repump beams are 42 mW and 2 mW, respectively. Using a magnetic field gradient of $\sim$24 G/cm at the center of the MOT,  typically 10$^7$ atoms are trapped, and the typical  Gaussian full-width-half-maximum (FWHM) of the MOT is 0.5 mm. The free-space MOT fluorescence is collected by an achromatic  doublet ($f=150$ mm) and is divided into two parts using a 50:50 beam-splitter. One half of the signal is collected by a PMT to measure the instantaneous number of atoms in the MOT. The other half is imaged by an EMCCD camera to obtain the atom cloud density profile. The MOT fluorescence that couples to the guided-mode of the ONF is separated from light of other wavelengths using an assembly of dichroic mirrors and bandpass filters, before being delivered to an SPCM. The overlap between the cold atom cloud and the ONF is optimized by maximizing the photon count at the SPCM using three pairs of Helmholtz coils. With an optimized overlap, the SPCM count is proportional to the atom density, and, hence, the PMT signal, for a given Gaussian FWHM. 

The Rydberg excitation is driven by a ladder-type, two-photon process, shown in Fig. \ref{Fig:setup}(b). The 780 nm light is provided by the cooling beams, while the 482 nm light is derived from a Toptica TA SHG Pro system and coupled into the nanofiber. The frequency of the 482 nm laser is stabilized to a vapor cell electromagnetically induced transparency (EIT) signal \cite{Adams2007, doi:10.1063/1.3086305}. However, we implement a  modification using an electro-optic modulator (EOM) to shift the frequency. First, the 780 nm probe laser is locked to the $5S_{1/2}(F=2)\rightarrow5P_{3/2}(F^{\prime}=2,3)_{co}$ transition of $^{85}$Rb. The EOM is used to generate sidebands at 1.06632 GHz, one of which is resonant with the $^{87}$Rb 5$S_{1/2}(F=2) \rightarrow$ 5$P_{3/2}(F^{\prime} = 3)$ transition. In the $^{87}$Rb-enriched vapor cell, only the resonant sideband participates in the EIT process and the resultant EIT peak is used to lock the 482~nm laser frequency, which can now be adjusted  simply by changing the frequency of the sideband.

The 482 nm light can be switched on and off using a combination of an AOM and a mechanical shutter to ensure complete cut-off. This light is coupled to the ONF using a pair of dichroic mirrors (DMLP650) and  interacts with the atoms in the MOT via the evanescent field. The coupling efficiencies for light of different wavelengths into the fiber patch cable differ, as do the losses at any splice regions, making it difficult to determine the exact transmission through the ONF for each wavelength used in the experiment.  We, therefore, measure the power at the output pigtail of the ONF in order to estimate the power at the nanofiber waist. Unless another power is explicitly mentioned, we maintain an output power of 16$~\mu$W for the 482 nm light for all measurements. Note that, at 482 nm, the ONF supports the fundamental mode, $HE_{11}$, as well as the $TE_{01}$, $TM_{01}$, and $HE_{21}$ higher order modes. The ONF is single-mode for all other wavelengths used.

The experimental sequence is simple. First, the MOT is loaded to saturation for 8 seconds. The population of the MOT at any time, $t$, as a fraction of the undisturbed equilibrium population, can be obtained from the PMT signal and is expressed as 

\begin{equation}
        N(t)=\frac{N_{tot}(t)}{N_{0}} = (1-e^{-\gamma_{m}t}),
\label{eq1}
 \end{equation}

 \noindent where $N_{tot}(t)$ is the total number of atoms in the MOT at time $t$ and $N_{0}$ = $L$/$\gamma_{m}$ is the steady-state number of atoms in the MOT.  $L$ is the loading rate of atoms into the MOT from the background vapor and $\gamma_{m}$ is the population loss rate of the MOT. A typical loading curve is shown in Fig. \ref{Fig:setup}(c) (left-hand side of the plot) and $\gamma_{m}$ is obtained by fitting the PMT signal to Eq. \ref{eq1}. 
 
 When the MOT is loaded to saturation, the 482 nm laser propagating in the ONF is turned on. Only those atoms in the evanescent field of the nanofiber can interact with both the 780 nm and 480 nm light and, therefore, participate in the two-photon Rydberg excitation. The newly formed  Rydberg atoms leave the cooling cycle and escape from the MOT - this introduces a new population loss rate, $\gamma_{R}$, which includes any other loss processes, such as ionization of atoms post Rydberg excitation. The new loss mechanism starts bleeding the MOT of atoms and a new equilibrium is established over time, determined by the total loss rate, $\gamma_{t} = \gamma_{m}+\gamma_{R}$. The time-dependent population of the MOT can be written as  

\begin{equation}
       N(t) = 1-\frac{\gamma_R}{\gamma_t}(1-e^{-\gamma_t(t-t_0)}),
\label{eq2}
 \end{equation}

\noindent where $t_0$ is the time at which the 482 nm laser is switched on. The time evolution of the MOT population is fitted to Eq. \ref{eq2} to obtain $\gamma_R$. Assuming all atoms excited to the Rydberg state are lost from the cooling cycle, $N_0\gamma_R$ is the rate of formation of the Rydberg atoms at the moment the 482 nm laser is switched on. 

\begin{figure}[t]
\center
\includegraphics[width=7.0 cm]{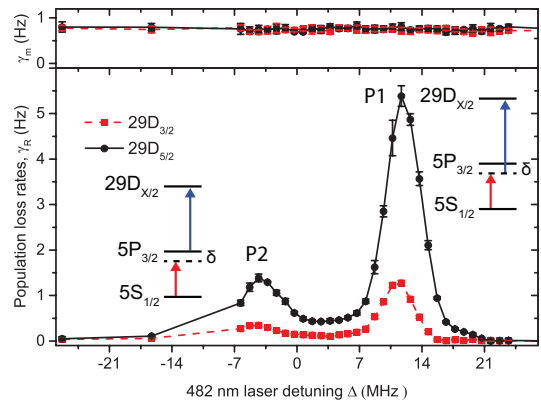}
\caption{Dependence of the population loss rates, $\gamma_m$ (top) and $\gamma_R$ (bottom), on the 482 nm laser detuning, $\Delta$, for a fixed detuning of $\delta = 14$ MHz for the 780 nm laser. For both levels, $29D_{3/2}$ and $29D_{5/2}$, $\gamma_m$ does not change while $\gamma_R$ shows two peaks. The peaks are labeled with the level schemes to indicate the coherent (P1) and the incoherent (P2) processes \cite{PhysRevA.14.802}.}
\label{Fig:signal4}
\end{figure}

For a given detuning, $\Delta$,  of the 482 nm laser from the $5P_{3/2}$ to the $29D$ Rydberg level, the experiment is repeated for 8 cycles to obtain mean values of $\gamma_m$ and $\gamma_R$. The variation of $\gamma_m$ and $\gamma_R$ as a function of $\Delta$ is investigated for two Rydberg levels, namely $29D_{5/2}$ and $29D_{3/2}$. The results are presented in Fig. \ref{Fig:signal4}. Note that $\gamma_m$ does not change during the experiment, ensuring that the experimental conditions also do not change.  We can clearly see that $\gamma_R$ shows two peaks as a function of $\Delta$ for both of the $29D$ levels. The two peaks, i.e., P1 at $\Delta=11.7$ MHz and P2 at $\Delta =-4.3$ MHz, correspond to two different mechanisms for the Rydberg excitation.  P1 is obtained from a coherent, two-photon excitation, where a fraction of the ground-state atom population is coherently transferred to the Rydberg state without populating the intermediate, $P_{3/2}$, excited state. In contrast, P2 corresponds to an incoherent, two-step excitation process. The cooling laser transfers a fraction of the ground-state population to the intermediate state, $P_{3/2}$. The second photon, at 482 nm, then transfers a fraction of the $P_{3/2}$ population to the Rydberg state. In this process, the intermediate state is populated and there is no coherence established between the ground and  Rydberg states. A detailed explanation of the mechanism affecting the ratio between coherent and incoherent excitations can be found in Ref. \cite{PhysRevA.14.802}. Ideally, the peaks should appear at $\Delta$ = 0 and $\delta$ (the detuning of the 780 nm cooling laser). The deviation of the peaks from the expected position and separation values may arise from many factors, such as a light shift, a van der Waal's shift due to the fiber surface \cite{Hakuta2004,Minogin2010, Frawley_2012, Patterson_2018}, or a shift due to stray electric charge on the fiber surface \cite{Windpassinger2017}, experienced by the energy levels involved in the excitation process. These effects have not been incorporated in the model presented herein.  

\begin{figure}[t]
\center
\includegraphics[width=7.0 cm]{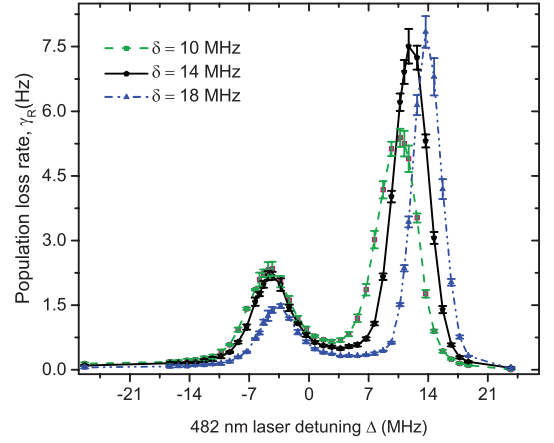} 
\caption{Dependence of the population loss rate, $\gamma_R$, on the 482 nm laser detuning, $\Delta$, for three different detunings of 780 nm, i.e, $\delta = 10$, $14$, and $18$ MHz, for the $29D_{5/2}$ level.}
\label{Fig:detuning}
\end{figure}

Figure \ref{Fig:detuning} shows the variation of $\gamma_R$ as a function of $\Delta$, for three different values of the cooling laser detuning, $\delta$. The position of the coherent peak, P1, changes with $\delta$ to satisfy the two-photon resonance condition.  In this process, the sum of the energies of the 780 nm photon and the 482 nm photon should be equal to the energy difference between the ground and Rydberg state. The $5P_{3/2}$ intermediate state is always populated following a small change in detuning of the cooling laser. Therefore, the peak position should not change since the 482~nm photon should always have the same energy in order to couple the $5P_{3/2}$ intermediate state to the Rydberg level; however, this argument ignores other effects that could lead to energy level shifts in the experiment. As we expect, the position of P2 does not noticeably change  for $\delta=10$ MHz and 14 MHz; however, there is a small, observable shift when $\delta=18$ MHz.  A model accounting for surface and charge effects in a more realistic geometry would be needed to determine the reason for this observed shift.

\begin{figure}[ht]
\center
\includegraphics[width=8.6 cm]{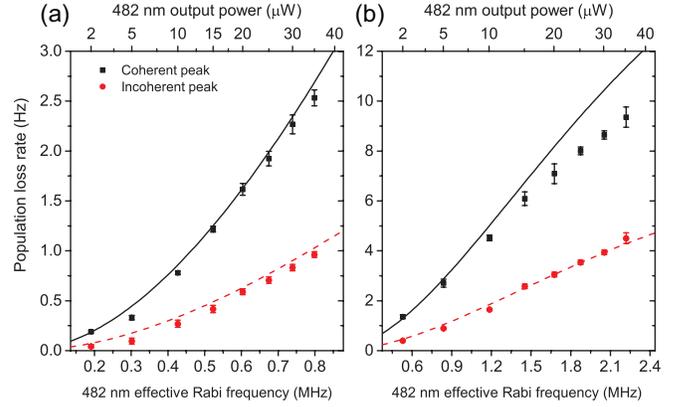}
\caption{The dependence of $\gamma_R$ for the coherent (P1 black) and incoherent (P2 red) excitation on the power of the 482 nm pump laser and, hence, the effective Rabi frequency for (a) $29D_{3/2}$ and (b) $29D_{5/2}$. The solid lines are solutions to the three-level model.}
\label{Fig:power}
\end{figure}

Finally, we investigated the dependence of $\gamma_R$ on the effective Rabi frequency of the $5P_{3/2} \rightarrow 29D_{X/2}$ transition for both the coherent and  incoherent processes. To perform this experiment, $\delta$ was set to 14 MHz and $\Delta$ was set to the maximum of either P1 or P2. The power of the 482 nm laser was then varied and $\gamma_R$ was measured for both  peaks and both  transitions. The results are shown in Fig. \ref{Fig:power}. We compare the dependence of the observed loss rates on the frequencies and intensities of the driving optical fields by considering a three-level density matrix model for a population of atoms driven by two coherent optical fields of constant intensity. We use the Maxwell-Bloch equations for thermal atoms in a small interaction volume. As the atoms interact with different intensities of the evanescent field at different distances from the ONF surface, the Rabi frequency is position dependent. We define an effective Rabi frequency, $\Omega_r$, averaged over the atom ensemble interacting with the evanescent field for the $5P_{3/2} \rightarrow 29D_{X/2}$ transition. Using the definitions of the Rabi frequency, $\Omega_p$, for the cooling transition, the effctive Rabi frequency, $\Omega_r$, for the Rydberg transition, detunings from resonance, $\Delta_p$ and $\Delta_r$, and atomic operators, $\sigma_{ij}$, representing populations and coherences of all three levels, we can write:
\begin{align*}
\partial_t\sigma_{ss}=&\left[i\frac{\Omega_p}{2}\sigma_{ps}+H.c.\right]+\Gamma_p\sigma_{pp}+\Gamma_r\sigma_{rr}\\&-A(\sigma_{ss}-\frac{1}{2}),\\
\partial_t\sigma_{pp}=&\left[i\left(\frac{\Omega_r}{2}\sigma_{rp}-\frac{\Omega_p}{2}\sigma_{ps}\right)+H.c.\right]-\Gamma_p\sigma_{pp}\\&-A(\sigma_{pp}-\frac{1}{2}),\\
\partial_t\sigma_{rr}=&\left[i\frac{\Omega_r}{2}\sigma_{rp}+H.c.\right]-\left(\Gamma_r+A\right)\sigma_{rr},\\
\partial_t\sigma_{ps}=&-i\Delta_p\sigma_{ps}+i\frac{\Omega_p}{2}(\sigma_ss-\sigma_pp)+i\frac{\Omega_r}{2}\sigma_{rs}\\&-(\frac{\Gamma_p}{2}+A)\sigma_{ps},\\
\partial_t\sigma_{rp}=&-i\Delta_r\sigma_{rp}+i\frac{\Omega_r}{2}(\sigma_{pp}-\sigma_{rr})-i\frac{\Omega_r}{2}\sigma_{rs}\\&-(\frac{\Gamma_p+\Gamma_r}{2}+A)\sigma_{rp},\\
\partial_t\sigma_{rg}=&-i(\Delta_p+\Delta_r)\sigma_{rp}+i\frac{\Omega_r}{2}\sigma_{ps}-i\frac{\Omega_p}{2}\sigma_{rp}\\&-(\frac{\Gamma_r}{2}+A+\gamma_0)\sigma_{rs}.
\end{align*}

\noindent Here, $\gamma_0$ is the motional dephasing of the coherence, $\sigma_{rg}$. For a MOT temperature of 120 ~$\mu$K and a coherence generated by a cooling beam perpendicular to the nanofiber axis,  $\gamma_0 = 600 \pm 200$ kHz. Other motional dephasings are ignored. The $+ H.c.$ terms indicate that the Hermitian conjugates of the terms in $[~]$ are also included. The thermal motion of atoms into and out of the interaction volume removes atoms from the populations and coherences at a rate $A$ and places them in a mixture of roughly $s\sigma_{ss}+p\sigma_{pp}$ with $s$ and $p$ determined by $\Omega_p$ and $\Delta_p$.
 The model ignores other aspects of the experimental geometry. $\Omega_p$ is fitted from the splitting of P1 and P2, and $\Omega_r$ is fitted to the P1 and P2 height data. The decoherence rates are fitted to the widths of P1 and P2.

The mixing rate, $A$, is an important parameter for describing this experiment. The introduction of atoms in a mixed state to the interaction region boosts the incoherent production of Rydberg atoms well beyond that expected in a coherently driven system. The mixing rate may be changed experimentally by changing the cooling field detuning and thus the temperature of the atoms. The effect of this can be seen in Fig. \ref{Fig:detuning}, where a closer detuned trap with hotter atoms has a higher mixing rate and a larger incoherent peak to coherent peak height ratio than a further detuned, cooler trap.
 
 The population of Rydberg atoms in the model, $\sigma_{rr}$, is related to the experimental loss rate, $\gamma_R$, by estimating the proportion, $P$, of atoms in the MOT that are also in the evanescent field and multiplying by the mixing rate: $PA\sigma_{rr}\approx \gamma_R$. The value of $P$ is set to $P=(3\pm1)*10^{-4}$ to fit the model  to the experimental data in Fig. \ref{Fig:power}, and this value also agrees with other experimental observations. It corresponds to an interaction region extending about 100 - 200 nm from the fiber surface, noting that the $1/e$ decay length of the  evanescent field of the fundamental mode of 482 nm light is 125 nm. The mixing rate, $A$, is set to 0.6 MHz to fit the ratio of the incoherent to coherent peak heights in Fig. \ref{Fig:power}. This value is also consistent with the average flight time of atoms at 120 $\mu$K through an interaction region with a diameter, $d=200$ nm, where $A$ $\approx \bar{v}/d$, with $\bar{v}$ the average speed of the atoms. 

The experimental loss rate clearly follows the theoretical curve at low $\gamma_R$, but diverges from the model above rates of 5 Hz. This may be due to production of Rydberg atoms beginning to saturate as $\sigma_{rr}\rightarrow 1$. This is not observed in the model, with a Rydberg population at P1 of $\sigma_{rr} = 0.033$ for $\Omega_r=1.25$ MHz, which is well below saturation. The model only includes one cooling field, with a Rabi frequency equal to that of one cooling beam. $\Omega_r$ is assumed to be constant, whereas, in the experiment, atoms are subject to a time-varying interaction as they travel through the evanescent field. Averaging over these trajectories may give a more accurate relationship between the 482 nm intensity and the effective $\Omega_r$. The model also assumes that all atoms leaving the interaction region while in the Rydberg state are lost from the MOT. However, considering the lifetime of the Rydberg states and the temperature of the atoms, a significant proportion of these atoms could actually be recaptured directly into the MOT if they decay into the cooling cycle. Experimental determination of the saturated loss rates under different conditions and for different $n$ levels could allow any processes interfering with recapture, such as ionization, to be observed. However, despite its simplicity, the model explains most of our observations and confirms the production of Rydberg atoms in the evanescent field of the nanofiber.

In summary, we have reported on the generation of \textit{n} = 29 level Rydberg atoms in a $^{87}$Rb cold atom ensemble surrounding an optical nanofiber and have shown that this is a very viable system for hybrid atom-nanofiber devices. By comparing the loss rate to the Rydberg population in a simple 3-level density matrix model, we were able to closely model the experimental data and determine the size of the interaction region.  Excitation of the Rydberg atoms is mediated via the optical nanofiber and they are estimated to be generated at no more than a couple of 100 nm from the surface where overlap between the two excitation fields (780 nm and 482 nm) is maximum. This is an important advance on earlier experiments related to Rydberg atom generation close to dielectric surfaces \cite{PhysRevLett.116.133201,PhysRevA.81.063411} and opens up many avenues of research such as all-fibered quantum networks using Rydberg atoms, excited atom-surface interactions at submicron distances, including van der Waal's interactions, effects on the Rydberg blockade or facilitation phenomena \cite{PhysRevLett.118.063606} in this new regime, stray electric field effects from the dielectric nanofiber on energy levels and lifetimes of the atom, and the limitation of the maximum excited state (\textit{n} value) that can be generated close to the nanofiber due to the atom size increasing quadratically with \textit{n}. An estimate of the electric field present and its influence on the Rydberg atom lifetime is highly desirable and will be the focus of the next generation of experiments. For now, we can assume that the charge per unit length of the tapered fiber is no more than 2 x 10$^{11}$ C/m \cite{Kamitani:16}.  This number could increase when light propagates through the fiber \cite{Harlander_2010}, but the polarizability of the 29$D$ state indicates that the field must be limited to only a few V/cm in order to generate the Rydberg atoms at the right laser detuning.  A systematic study related to the Rydberg atom formation and lifetime as a function of power in the blue beam is also desirable.  The observed red-shifts on the resonance frequencies are assumed to be dominated by van der Waals interactions between the atoms and the fiber surface \cite{Minogin2010, Frawley_2012, Patterson_2018}.

The versatility of this atom-nanofiber hybrid system could be extended to explore three-step Rydberg excitations \cite{Fahey:11} where the fiber would be single-mode for the light used to drive the atomic transitions. Hence, the mode overlap in the evanescent field would be increased and Rydberg generation efficiency should be improved.  A loss in detection of the 420 nm light in the $6P \rightarrow 5S$ decay channel could also provide an alternative, non-destructive mechanism for Rydberg atom detection.  In addition, a comprehensive study of the coherent interactions in MOT-embedded nanofibers could extend this experimental technique beyond a qualitative confirmation towards an investigation of the behavior of Rydberg or other exotic states, e.g. Rydberg polarons \cite{PhysRevLett.120.083401}, adjacent to  optical nanofibers.  Future work will focus on trapping atoms at defined distances from the nanofiber to explore limitations on \textit{n} and on a study of the influence of the nanofiber on Rydberg levels using EIT signals \cite{Adams2007}. 

This work was  supported by OIST Graduate University, JSPS Postdoctoral Fellowship for Overseas Researchers Grant Number JPPE16769, and JSPS Grant-in-Aid for Scientific Research (C) Grant Number 19K05316.   \\
The authors acknowledge early contributions by M. Langbecker and useful discussions with E. Stourm, J. Robert, E. Brion, and R. L\"ow.
K.S.R., T.R. and K.K. conducted the experiments. J.E. did the theoretical modelling. S.N.C. and T.R. conceptualized the work. All authors wrote the manuscript.

%

\end{document}